\documentclass[a4paper,fleqn,12pt]{article} % default [10pt]

\usepackage{fullpage} % for article style A4

\usepackage{multirow} % for TLP
\usepackage{graphicx} % for TLP

\usepackage{url}
\usepackage{alltt}
\usepackage{amsmath} % for align and gather for line breaks in math mode, equations, % does not really work
\usepackage{amssymb}

\usepackage{epsfig}
\usepackage{wrapfig}
\usepackage{subfigure}

\usepackage{rotating}
\usepackage{moreverb}
\usepackage{fancyvrb}

\def\bd{\begin{description}}
\def\ed{\end{description}}
\def\bc{\begin{center}}
\def\ec{\end{center}}
\def\bq{\begin{quote}}
\def\eq{\end{quote}}
\def\bi{\begin{itemize}}
\def\ei{\end{itemize}}
\def\be{\begin{enumerate}}
\def\ee{\end{enumerate}}
\def\ba{\begin{array}}
\def\ea{\end{array}}

\newcommand{\true}{{\mathit true}}

\newcommand \aeq {:=}  % arithmetic equality, was :=
\newcommand{\false}{{\mathit false}}

\newtheorem{definition}{Definition}[section]
\newtheorem{example}{Example}[section]

\newcommand{\simp}{{\; \leftarrow \;}}

\newcommand{\myparagraph}[1]{\textbf{#1.}}

\begin{document}

\title{Super-Linear Speedup by Generalizing Runtime Repeated Recursion Unfolding in Prolog}

\author{%
Thom Fr{\"u}hwirth\\
Ulm University, 89069 Ulm, Germany\\
thom.fruehwirth{@}uni-ulm.de%
}

\maketitle

\begin{abstract}

Runtime repeated recursion unfolding was recently introduced in \cite{recfundam} as a just-in-time program transformation strategy that can achieve super-linear speedup.
So far, the method was restricted to single linear direct recursive rules
in the programming language Constraint Handling Rules (CHR). 
In this companion paper, we generalize the technique to multiple recursion and to multiple recursive rules and provide an implementation of the generalized method in the logic programming language Prolog.

The basic idea of the approach is as follows:
When a recursive call is encountered at runtime, the recursive rule is unfolded with itself and this process is repeated with each resulting unfolded rule as long as it is applicable to the current call. In this way, more and more recursive steps are combined into one recursive step. Then an interpreter applies these rules to the call starting from the most unfolded rule.
For recursions which have sufficiently simplifyable unfoldings, a super-linear can be achieved, i.e. the time complexity is reduced.

We implement an unfolder, a generalized meta-interpreter and a novel round-robin rule processor for our generalization of runtime repeated recursion unfolding with just ten clauses in Prolog.
We illustrate the feasibility of our technique with 
worst-case time complexity estimates and benchmarks for some basic classical algorithms
that achieve a super-linear speedup. 
\end{abstract}


\tableofcontents

\section{Introduction}\label{intro}

In this companion paper, we generalize our online program optimization strategy of runtime repeated recursion unfolding from the main paper \cite{recfundam} to enable super-linear speedups for more classes of recursion.
We generalize the approach to multiple recursion and to multiple recursive rules and we provide an implementation of the generalized method in the logic programming language Prolog.

Unfolding is a program transformation that basically replaces a call in the body (right-hand side) of a rule with the body of a rule whose head (left-hand side) fits the call. 
\emph{Runtime Repeated recursion unfolding} \cite{reclopstr,recfundam}
first unfolds a given recursive rule with itself and simplifies it. 
This results in a specialized recursive rule that, when applied, covers two combined recursive steps instead of one.
The technique continues to unfold the last unfolded recursive rule with itself
until the resulting rule is no longer applicable to the given call.
Each unfolding doubles the number of recursive steps covered.
Then, the rules are applied to the given recursive call, with the most unfolded rule being tried first, continuing until the original rule and finally a base case are reached.
If the unfolded rules admitted sufficient simplification of the combined recursive steps, this approach results in a proven super-linear speedup \cite{recfundam}.

In this paper, we assume some familiarity with Prolog \cite{sterling1994art}.
For efficiency, we assume that each recursive clause has a {\em green cut}. 
To ease the explanation of the method in this paper, we use the term {\em guard} for the goals before the cut and the term {\em body} only for the goals after the cut.
\begin{example}[Summation]
Consider the following introductory example adapted 
from \cite{recfundam}, a simple recursive program
written here in the abstract syntax of Prolog. 
It recursively adds all numbers from $1$ to $n$.
Rule $b$ covers the base case and rule $r$ covers the recursive case.
The {\em head} $sum(N,S)$, {\em guard} (e.g. $N{=}1$) and {\em body} of a rule are separated by the 
symbols $\simp$ and cut $!$, respectively.
\begin{gather*}
  b= sum(N,S) \simp N=1 ,!, S=1\\
  r= sum(N,S) \simp N>1 ,!, sum(N{-}1,S1), S \aeq N{+}S1
\end{gather*}
Unfolding the recursive rule with a copy of itself and simplifying the resulting rule gives
$$r_1= sum(N,S) \simp N>2 ,!, sum(N{-}2,S1'), S \aeq 2{*}N{-}1{+}S1'$$
Note that this rule $r_1$ behaves like applying the original rule $r$ twice. 
With rule $r_1$ we only need about half as many recursive steps as with the original rule alone. 
Because the arithmetic computation is simplified, we can also expect to halve the runtime.

We proceed with unfolding rule $r_1$ with itself:
$$r_2= sum(N,S) \simp N>4 ,!, sum(N{-}4,S1), S \aeq 4*N{-}6+S1$$
This rule results in fourfold speedup.
We can continue this process, doubling the speed each time\footnote{Clearly there is a closed-form solution for this problem, $S=N*(N+1)/2$, but this is not the point of the example.}.
This kind of simplification of the combined recursive steps in the unfolded rules is sufficient to achieve a super-linear speedup.
The number of unfoldings depends on the given call. 
The most unfolded rule should cover as many recursive steps of the call as possible.
For example, 
for $N{=}4$ we will unfold till rule $r_1$ with guard $N{>}2$,
for $N{=}5$ till rule $r_2$ with $N{>}4$,
for $N{=}50$ till rule $r_5$ with $N{>}32$.
\end{example}

Repeated recursion unfolding requires unfolding on-the-fly because the number of unfoldings depends on the current call. 
We do not want to modify the given program at runtime.
Therefore, the method also introduces an interpreter for the unfolded rules.
This \emph{meta-interpreter\footnote{A meta-interpreter interprets a program written in its own implementation language (Chapter 8 in \cite{huntbach1999agent}).}} tries and applies each unfolded rule at most once starting with the given call and the most unfolded rule.

\myparagraph{Overview of the Paper}
As a companion paper, this work is not fully self-contained. To make it more accessible, we reuse some parts from the main paper \cite{recfundam}.

After this introduction,
\emph{Section \ref{rrrun}} defines the program transformation method of runtime repeated recursion unfolding with simplification
and an optimal rule application strategy 
for the logic programming language Prolog.
The exposition follows the one in the main paper, but replaces CHR with Prolog.

\emph{Section \ref{impl}} presents our lean implementation of the unfolder and meta-interpreter to perform repeated recursion unfolding with optimal rule applications at runtime, 
again following the main paper. 
\emph{Section \ref{genimpl}} generalizes our implementation of runtime repeated recursion unfolding for multiple recursive rules using a new program layer called the round-robin rule processor.
This is the main contribution of this companion paper.
The complete implementation consists of just ten clauses written in Prolog.
It has little overhead, the runtime mainly depends on the given recursive rule and its unfoldings.

\emph{Section \ref{secbench}} contains the experimental evaluation of our technique on some examples, summation, Fibonacci numbers and greatest common divisor (GCD). 
Fibonacci features multiple recursion and GCD features multiple recursive rules.
We derive the necessary simplifications, discuss the time complexity and compare it with the result of benchmarks to verify the super-linear speedup.

\emph{Section \ref{related}} discusses related work.
Finally, the conclusion discusses achievements, limitations and future work of generalized runtime repeated recursion unfolding. 
In the appendix we give two additional examples, list reversal and sorting from \cite{recfundam}.

\section{Repeated Recursion Unfolding in Prolog}\label{rrrun}

We give a short definition of the semantics of Prolog 
and of rule unfolding in Prolog \cite{sterling1994art}.
Then we introduce repeated recursion unfolding based on \cite{recfundam}
and define an optimal rule application strategy.

\subsection{Semantics of Prolog}%\label{prelim}

Logic Programming \cite{sterling1994art,KowalskiBook} is a programming and knowledge representation paradigm based on first-order logic. 
We assume some familiarity with the syntax and semantics of the logic programming language Prolog:
{\em Goals} are possibly empty conjunctions of atoms.
They are denoted by upper case letters in definitions. 
To avoid clutter, we use simple commas to denote logical conjunction.
Prolog built-in predicates include $\true$ and $\false$,  
the syntactic equality $=/2$ between terms and arithmetic operations using the arithmetic equality $\aeq/2$.
Syntactic equality tries to unify its arguments, i.e. making them syntactically identical by instantiating their variables appropriately.

To define the operational semantics of Prolog,
we assume a transition system with a transition relation $\mapsto$ between states that are goals.
An initial state is any goal, a final state is one that contains only syntactic equalities (including the special cases $\true$ and $\false$).
\begin{definition}[Prolog Semantics]\label{def:prolog}
{%\rm
A Prolog program ${\cal P}$ consists of clauses of the form
$H \leftarrow B$, where $H$ is an atom and $B$ is a goal.
If $B$ is empty, it is equivalent to $\true$ and the clause is called a {\em fact}, otherwise it is called a {\em rule}.
Let $\cal S$ be a function that simplifies the syntactic equalities in a goal.

Given a goal, let $G$ be an atom chosen from the goal and $D$ be the remainder of the goal.
\begin{itemize}
\item Choose a clause from ${\cal P}$ and take a copy of the clause with new variables, $H \leftarrow B$.
If $G$ and $H$ are unifiable, then
$$G, D \mapsto {\cal S}(G{=}H, B, D).$$

\item Otherwise, if $G$ is a call to a built-in predicate, the transition replaces it by the result of executing the built-in, where the result is a conjunction of syntactic equalities $E$,
$$G, D \mapsto {\cal S}(E, D).$$

\item Otherwise, $G$ gives rise to a failure transition into the failed state denoted by $\false$,
$$G, D \mapsto \false.$$
\end{itemize}%
} %rm
\end{definition}
In implementations of Prolog, all clauses are tried for the current goal in a systematic manner using chronological backtracking. In the context of this paper, we assume green cuts for efficiency  \cite{sterling1994art} in the recursive clauses.
This means that we commit to the chosen clause once the cut is reached.

\subsection{Rule Unfolding with Simplification}

We now define unfolding with simplification for Prolog rules.
\begin{definition}[Unfolding]\label{def:unf}
{%\rm
Let ${\cal P}$ be a Prolog program and let $r$ and $v$ be 
copies of two (not necessarily different) clauses from ${\cal P}$ such that they do not share variables
$$
\begin{array}{lcl}
r= H & \leftarrow &  D, G, B\\
v= H' & \leftarrow & B',
\end{array}
$$
where $H, H'$ and $G$ are atoms,
and $D, B$ and $B'$ are goals. 

The \emph{unfolding of rule $r$ with rule $v$, $\mathit{unfold}(r,v)$,} is
the rule
$$r'= H \leftarrow D, G{=}H', B', B \ \mbox{  if $G$ and $H'$ are unifiable}.$$ 
} %rm
\end{definition}
If the goal $G$ in the body of rule $r$ unifies with the head $H'$ of a rule $v$,
unfolding replaces $G$ by $G{=}H'$ and by the body of rule $v$ to obtain the unfolded rule $r'$.

In the presence of a cut operator in the body of a rule, the correct unfolding in general becomes more complicated \cite{Prestwich1993}. 
Here we assume green cuts in the recursive rules.
In this case, it suffices to remove the cut coming from the rule $r$ in the unfolded rule $r'$.
The cut from rule $v$ will stay.

Speedup crucially depends on the amount of {\em simplification} of the combined recursive steps in the unfolded rules. 
We want to replace goals in the recursive step of the rules by semantically equivalent ones that can be executed more efficiently.
This includes moving the green cut to the left as far as semantically possible.
For simplification, we can use any available program transformation technique.

\begin{example}[Summation, contd.]\label{sumunf}{%\rm
We unfold the recursive rule for summation with (a copy of) itself:
\begin{gather*}
r= sum(N,S) \simp N>1 ,!, sum(N{-}1,S1), S \aeq N{+}S1\\
v= sum(N',S') \simp N'>1 ,!, sum(N'{-}1,S1'), S' \aeq N'{+}S1'
\end{gather*}
Then the unfolded rule is
\begin{gather*}
r_1= sum(N,S) \simp N>1, sum(N{-}1,S1){=}sum(N',S'), N'>1 ,!,\\
\hspace{0.9cm} sum(N'{-}1,S1'), S' \aeq N'{+}S1', S \aeq N{+}S1
\end{gather*}
The unfolded rule can be simplified into the rule
$$r_1= sum(N,S) \simp N{>}2 ,!, sum(N{-}2,S1'), S{\aeq}2{*}N{-}1{+}S1'$$
} %rm
\end{example}

\subsection{Repeated Recursion Unfolding}

We can now define the program optimization strategy of repeated recursion unfolding for Prolog based on rule unfolding with simplification. 
In our method, we start from a call (query) for a Prolog predicate defined by a recursive rule.
We unfold the recursive rule with itself and simplify it. 
Then we unfold the resulting rule.
We repeat this process as long as the resulting rules are applicable to the query.
\begin{definition}[Repeated Recursion Unfolding]\label{def:rru}
{%\rm
Let $r$ be a recursive rule and $G$ be a goal.  
Let $$\mathit{unfold}(r) = \mathit{unfold}(r,r).$$
The {\em repeated recursion unfolding} of a recursive rule $r$ with goal $G$ and with rule simplification
is a maximal sequence of rules $r_0, r_1, \ldots$ 
where rule $r_i$ has head $H_i$ and guard $C_i$ and where
\begin{gather*}
r_0 = r\\ 
r_{i+1} = \mathit{unfold}(r_i) \ \mbox{  as long as }G{=}H_i, C_i \not\mapsto \false \ (i\geq 0)%\\
\end{gather*}
} %rm
\end{definition}
We unfold and simplify the current unfolded rule $r_i$
if the guard of the rule 
is successfully executable with the query $G$. 
We add the new rule to the sequence and continue with it. 
Note that unfolding may fail to produce a rule. In that case, the unfolding process stops.

\begin{example}[Summation, contd.]{%\rm
Consider a query $sum(10,R)$.
Recall the unfolded simplified rule
$$r_1 = sum(N,S) \simp N{>}2 ,!, sum(N{-}2,S1), S \aeq 2{*}N{-}1{+}S1$$
The query $sum(10,R)$ means that the guard $N{>}2$ succeeds since $10=N$,
so we repeat the unfolding:
\begin{gather*}
\mathit{unfold}(r_1) = sum(N,S) \simp N{>}2,
sum(N{-}2,S1){=}sum(N',S'), N'{>}2 ,!,\\
\hspace{3cm}sum(N'{-}2,S1'), S' \aeq 2{*}N'{-}1{+}S1', S \aeq 2{*}N{-}1{+}S1
\end{gather*}
The unfolded rule can be simplified into the rule
$$r_2 = sum(N,S) \simp N{>}4 ,!, sum(N{-}4,S1'), S \aeq 4{*}N{-}6{+}S1'$$
The rule $r_2$ is applicable to the goal.
Further recursion unfolding results in rules with guards $N>8$ and then $N>16$.
To the latter rule, the goal $sum(10,R)$ is not applicable anymore. 
Hence repeated recursion unfolding stops.
The rules for the goal $sum(10,R)$ are therefore (more unfolded rules come first):
\begin{gather*}
r_3 = sum(N,S) \simp N>8 ,!, sum(N{-}8,S1), S \aeq 8*N{-}28+S1\\
r_2 = sum(N,S) \simp N>4 ,!, sum(N{-}4,S1), S \aeq 4*N{-}6+S1\\
r_1 = sum(N,S) \simp N>2 ,!, sum(N{-}2,S1), S \aeq 2*N{-}1+S1\\
r = r_0 = sum(N,S) \simp N>1 ,!, sum(N{-}1,S1), S \aeq N+S1\\
b = sum(N,S) \simp N=1 ,!, S=1.
\end{gather*}
Note that to the goal $sum(10,R)$ we can apply any of the recursive rules. The most efficient way is to start with the first, most unfolded rule. It covers more recursive steps of the original recursive rule than any other rule.
} %rm
\end{example}

\subsection{Optimal Rule Applications}

An unfolded rule covers twice as many recursion steps than the given rule.
When we apply a more unfolded rule, we cover more recursive steps with a single rule application.
Based on this observation we introduce a rule application strategy where we try 
to apply more unfolded rules first. Furthermore each unfolded rule is tried only once and is applied at most once.
In \cite{recfundam}, we proved this optimal rule application strategy sound and complete in the case of CHR. Here we apply it to Prolog.

\begin{example}[Summation, contd.]{%\rm
A computation with optimal rule applications for the goal $sum(10,R)$ is
as follows (where we subscript single transitions with the rule that was used):
\begin{gather*}
\hspace{-1cm}sum(10,R) \mapsto_{r_3}\\ 
\hspace{-1cm}{\cal S}(sum(10,R){=}sum(N,S), N>8, sum(N{-}8,S1), S \aeq 8*N{-}28+S1) \mapsto^+\\
\hspace{-1cm}sum(2,S1), R \aeq 52+S1 \mapsto_{r_0}\\
\hspace{-1cm}{\cal S}(sum(2,S1){=}sum(N',S'), N'>1, sum(N'{-}1,S1'), S' \aeq N'+S1', R \aeq 52+S1) \mapsto^+\\
\hspace{-1cm}sum(1,S1'), R \aeq 54+S1' \mapsto_b\\
\hspace{-1cm}R = 55
\end{gather*}
} %rm
\end{example}

\section{Implementation of Runtime Repeated Recursion Unfolding in Prolog}\label{impl}

At compile-time, the rules for the given {\em recursive predicate} are replaced by a call to the unfolder that contains these rules and then to the meta-interpreter that interprets the unfolded rules.
At runtime, the unfolder repeatedly unfolds a recursive rule as long as it is applicable to a given goal using a predefined {\em unfolding scheme} that specifies the simplification of the rule.
Then the meta-interpreter applies the resulting unfolded simplified rules according to the optimal rule application strategy.

In the following code in concrete Prolog syntax, 
{\tt =/2}, {\tt is/2}, {\tt copy\_term/2} and {\tt call/1} are standard built-in predicates of Prolog. 
The syntactic equality {\tt =/2} tries to unify its arguments, i.e. making them syntactically identical by instantiating their variables appropriately.
The arithmetic equality {\tt is/2} tries to unify its first argument with the result of evaluating the arithmetic expression in its second argument.
The built-in {\tt copy\_term/2} produces a copy of the given term with new variables. 
The Prolog meta-call {\tt call/1} executes its argument as a goal. 

Our implementation in SWI Prolog \cite{wielemaker2012swi} together with the examples is fully listed in this paper. Together with the benchmarking code it is also available from the author on request.

\subsection{Unfolder Implementation}\label{unfimpl}

The unfolder is implemented as a recursive Prolog predicate {\tt unf/3}.
It is the result of a straightforward translation from the unfolder written in CHR in \cite{recfundam}. 
It repeatedly unfolds and simplifies a recursive rule as long as it is applicable to a goal.
In {\tt unf(G,Rs,URs)}, the first argument {\tt G} is the goal and
{\tt Rs} is a list of given original clauses. 
{\tt URs} is the resulting list of unfolded and original clauses.

We assume that in the goal {\tt G} the input arguments are given and the output arguments are variables.
Initially, the list {\tt Rs} consists of a recursive rule followed by one or more clauses for the base cases of the recursion.
To simplify the implementation, the body of the rules in the lists
syntactically always consists of three conjuncts of goals:
the atoms before the recursive goal, the recursive goal and the atoms after the recursive goal. 
We use the built-in {\tt true} to denote an empty conjunct. 
\begin{verbatim}
 % unf(+RecursiveGoal, +RuleList, -UnfoldedRuleList)

unf(G, [R|Rs], URs) :-           % recursive step
  R = (H :- Co ,!, _),           % get head H and guard Co of rule R
  copy_term((H :- Co),(G :- C)), % copy them, unify head copy with goal G
  call(C)     	                  % call instantiated guard copy C
  ,!, 
  simp_unf(R, UR),               % unfold, simplify rule R into rule UR
  unf(G, [UR,R|Rs], URs).        % add new rule UR and recurse

unf(_G, [_R|Rs], URs) :- URs=Rs. % otherwise return rules Rs in URs
\end{verbatim}
In the recursive clause for {\tt unf/3},
we check if the rule {\tt R} in the list is applicable to the query (call, goal) {\tt G}:
If the guard {\tt C} succeeds, 
we unfold the current rule {\tt R} with itself and and simplify it using the predicate {\tt simp\_unf/2} (discussed below) and add the resulting rule {\tt UR} to the rule list in the recursive call of {\tt unf/3}.
When the guard has failed, the base case clause of {\tt unf/3} returns the rules that have been accumulated in the rule list as the result list in the third argument (with the exception of the first rule to which the goal was not applicable).

The Prolog predicate {\tt simp\_unf(R,UR)} 
takes the current rule and computes its simplified unfolding.
It is defined for each recursive rule.
In the head of a clause for {\tt simp\_unf/2} we use {\em rule templates} for the arguments to ease the implementation. 
The rules are then instances of the template.
Certain variables
in the template represent the {\em parameters} for the instance. 
These parameters will be bound at runtime.
In the body of a clause for {\tt simp\_unf/2}, the parameters for the unfolded rule will be computed from the parameters of the current rule.

The following example clarifies the above remarks on the implementation.
\begin{example}[Summation, contd.]\label{sumimpl}{%\rm
We show how we implement {\tt simp\_unf/2}
for the summation example. 
We abbreviate $sum$ to its first letter {\tt s} to avoid clutter in the code. 
The rule template for $sum$ is
\begin{verbatim}
s(A,C) :- A>V ,!, B is A-V, s(B,D), C is V*A-W+D % rule template
\end{verbatim}
where the variables {\tt V} and {\tt W} are parameters that stand for integers.
Its instance for the original recursive rule is
\begin{verbatim}
s(A,C) :- A>1 ,!, B is A-1, s(B,D), C is 1*A-0+D % rule instance V=1, W=0
\end{verbatim}
The implementation of the unfolding scheme for summation is accomplished by the following Prolog clause for {\tt simp\_unf/2}.
\begin{verbatim}
simp_unf(
  (s(A,C) :- A>V ,!, B is A-V, s(B,D), C is V*A-W+D), 
  (s(Al,Cl) :- Al>Vl ,!, Bl is Al-Vl, s(Bl,Dl), Cl is Vl*Al-Wl+Dl)
  ) :-
    Vl is 2*V, Wl is 2*W+V*V.
\end{verbatim}

For a goal {\tt s(100,S)} the unfolder is called with 
\begin{verbatim}
unf(s(100,S), [
     (s(A,C):-A>1 ,!, B is A-1, s(B,D), C is 1*A-0+D), % orig. recursion
     (s(A,B):-A=1 ,!, B=1, true, true)                 % base case
     ], URs). 
\end{verbatim}
It will return the following rules in the list {\tt URs}:
\begin{verbatim}
s(A,C) :- A>64 ,!, B is A-64, s(B, D), C is 64*A-2016+D
s(A,C) :- A>32 ,!, B is A-32, s(B, D), C is 32*A-496+D
s(A,C) :- A>16 ,!, B is A-16, s(B, D), C is 16*A-120+D
s(A,C) :- A>8 ,!, B is A-8, s(B, D), C is 8*A-28+D
s(A,C) :- A>4 ,!, B is A-4, s(B, D), C is 4*A-6+D
s(A,C) :- A>2 ,!, B is A-2, s(B, D), C is 2*A-1+D
s(A,C) :- A>1 ,!, B is A-1, s(B, D), C is 1*A-0+D     % orig. recursion
s(A,C) :- A=1 ,!, C=1, true, true                     % base case
\end{verbatim}
} %rm
\end{example}

\subsection{Generalized Meta-Interpreter Implementation for Multiple Recursion}

We implement the optimal rule application strategy with the help of a meta-interpreter for Prolog.
Again it closely follows the implementation for CHR in \cite{recfundam}. 
The meta-interpreter handles the recursive calls, any other goal will be handled by the underlying Prolog implementation. 
To a recursive goal, the meta-interpreter tries to apply the unfolded rules produced by the unfolder and applies each of them at most once.
The meta-interpreter is called with {\tt mip(G,Rs)}, where {\tt G} is the given recursive goal and {\tt Rs} is the list of rules from the unfolder {\tt unf/3}.
\begin{verbatim}
 % Meta-Interpreter for Multiple Recursion
 % mip(+RecursiveGoal, +RuleList)

mip(true,_Rs).         % base case, no more recursive goal

mip((G1,G2),Rs) :-     % handle conjunction for multiple recursion
   mip(G1,Rs),
   mip(G2,Rs).
mip(G,[R|Rs]) :-       % current rule is applicable to goal G
   copy_term(R, (G :- C ,!, B,G1,D)), % copy rule, unify head copy with G
   call(C)             % check guard
   ,!, 
   call(B),            % execute atoms before recursive call
   mip(G1,Rs),         % recurse with recursive goal and remaining rules
   call(D).            % execute atoms after recursive call
mip(G,[_R|Rs]) :-      % current rule is not applicable 
    mip(G,Rs).         % try remaining rules on G
\end{verbatim}
We now discuss the four clauses of our meta-interpreter.
In the first clause, the base case is reached since the recursive goal has been reduced to {\tt true}.
The second clause handles a conjunction of recursive goals by interpreting each conjunct on its own with the rule list {\tt Rs}. 
This case can handle multiple recursion and is new, it does not appear in the main paper \cite{recfundam}.

The third clause 
tries to apply the rule {\tt R} in the rule list to the current goal {\tt G}.
If the guard {\tt C} holds, the rule is applied.
The conjunct before the recursive goal {\tt B} is directly executed with a meta-call.
Next, the recursive goals {\tt G1} are handled with a recursive call to the meta-interpreter using the remainder of the rule list.
Finally the conjunct after the recursive goal {\tt D} is directly executed with a meta-call.

Otherwise the first rule from the rule list was not applicable, and so the last meta-interpreter clause recursively continues with the remaining rules in the list.

\subsection{Recursive Predicate Implementation}

In order to enable runtime repeated recursion unfolding,
at compile-time, the clauses for the given recursive predicate {\tt c/k} are replaced by a call to the unfolder {\tt unf/3} that contains these clauses and then to the meta-interpreter {\tt mip/2} that interprets the unfolded rules.
The replacement fits the following rule template where 
{\tt X1,...,Xk} are different variables.
\begin{verbatim}
 % rule template for a recursive predicate c/k

c(X1,...,Xk) :- 
     unf(c(X1,...,Xk), OriginalRules, UnfoldedRules),
     mip(c(X1,...,Xk), UnfoldedRules).
\end{verbatim}

\begin{example}[Summation, contd.]
For the summation example, the corresponding rule instance is as follows:
\begin{verbatim}
sum(N,S) :- 
    unf(s(N,S), [
       (s(A,C) :- A>1 ,!, B is A-1, s(B, D), C is 1*A-0+D),
       (s(A,C) :- A=1 ,!, C=1, true, true)
                 ], URs), 
    mip(s(N,S), URs).
\end{verbatim}
\end{example}

\section{Generalization to Multiple Recursive Rules}\label{genimpl}

We handle multiple recursive rules by handling each of the rules in separation. 
This is the main contribution of this companion paper.
Each recursive rule is subjected to runtime repeated recursion unfolding in a round-robin manner by a novel round-robin rule processor.

\subsection{Meta-Interpreter with Continuation Goals}

To enable this approach, we first generalize the original meta-interpreter by adding an additional argument that will contain the goal that remains after trying and applying the unfolded rules of one specific original recursive rule, we call it the {\em continuation goal}.

\begin{verbatim}
 % Meta-Interpreter with Continuation Goal
 % mip(+RecursiveCall, +RuleList, -ContinuationGoal)

mip(true,_LR,true) :- !.       % base case, no more goal

mip((R1,R2),LR,(G1,G2)) :- !,  % for multiple recursion
   mip(R1,LR,G1),
   mip(R2,LR,G2).
mip(R,[RR|LR],G) :-            % apply current rule
   copy_term(RR, (R :- C ,!, B,Rl,RCl)),  
   call(C)
   ,!,                
   call(B),
   mip(Rl,LR,G),
   call(RCl).  
mip(R,[_RR|LR],G) :-  % current rule not applicable
   mip(R,LR,G).

mip(R,[],R).          % no more rule applicable, return current goal
\end{verbatim}
The last clause is new, it is the additional base case {\tt mip(R,[],R)}.
It applies when the rule list has become empty. Then
the remaining unprocessed goal {\tt R} in the first argument is 
returned as continuation goal in the third argument.

\subsection{Round-Robin Rule Processor}

To handle the remaining goal, we then continue with generating, trying and applying the unfolded rules stemming from another recursive rule.
We repeat this process with all recursive rules until no goal remains.
This behavior is implemented by a {\em round-robin rule processor}
with a recursive predicate {\tt umr(R,RS)}, where {\tt R} is a recursive call and {\tt RS} is a list of lists of unfolded rules.
We keep and reuse unfolded rules between rounds.
In the comments in the code below, {\tt rl} stands for rule list
and {\tt rls} for rule lists.
\begin{verbatim}
 % Round-Robin Rule Processor
 % umr(+RecursiveCall, +RuleLists)
 % unfolding and interpretation rounds with several unfolded rule lists

umr(true,_RS) :- !.       % base case, no more recursive call to process

umr(R,[goal(G)|RS]) :- !, % this clause ensures termination
   R \= G,                % if goals unchanged, no rule was applied
   append(RS,[goal(R)],RS1), % add current goal at end of remaining rls
   umr(R,RS).             % continue with goal and rls

umr(R,[RL|RS]) :-  
   unf(R,RL,RL1),         % unfold with current rl, return extended rl
   mip(R,RL1,G),          % interpret with this rl, return contin. goal
   append(RS,[RL1],RS1),  % add expanded rl at end of remaining rls
   umr(G,RS1).            % recurse with contin. goal and updated rls
\end{verbatim}
In the first clause, the base case is reached since the goal is empty (which is represented by {\tt true}).

The second clause ensures termination in case no rule was applicable. The list element 
{\tt goal(G)} is reached after each round and holds the goal from the start of the round. When it is still equivalent to the current goal {\tt R}, no rule was applied to the goal and the computation fails due to the requirement {\tt R \verb+\=+ G}.

In the last, main clause, 
{\tt umr/2} tries to further unfold rules in the current rule list {\tt RL} with respect to the current recursive call {\tt R} using the unfolder {\tt unf/3} and then interprets {\tt R} with that rule list using {\tt mip/3}.
We add the extended rule list {\tt RL1} at the end of the rule lists {\tt RS} 
so that we can reuse it and can
go cyclically through all rule lists in a fair manner.
In the recursive call, the resulting continuation goal {\tt G} is further processed together with the updated rule lists {\tt RS1}.

\subsection{Recursive Predicate Implementation}

At compile-time, the rules for the given recursive predicate {\tt c/k} are replaced by a call to {\tt umr/2}, where 
{\tt X1,...,Xk} are different variables and
{\tt OriginalRuleLists} is 
the list of lists of the given recursive rules of {\tt c/k}.
We also add the clauses for the base cases at the end of each recursive rule list.
Finally, add the end of the lists of lists we add the list element {\tt goal(c(X1,...,Xk))} to ensure termination by recording the current call.
\begin{verbatim}
 % rule template for a recursive predicate c/k with multiple recursion

c(X1,...,Xk) :- umr(c(X1,...,Xk), OriginalRuleLists).
\end{verbatim}

\section{Experimental Evaluation with Benchmarks}\label{secbench}

Our examples will demonstrate that super-linear speedups are indeed possible. 
With sufficient simplification, the time complexity is effectively reduced when applying our generalized runtime repeated recursion unfolding.

For the worst-case time complexity of our implementation of runtime repeated recursion unfolding,
we have to consider the recursions in the original rules of the given program, in the unfolder, generalized meta-interpreter and round-robin rule processor.
We parametrize the time complexity by the number of recursive steps with the original rule. 
From the time complexity of the recursive steps we can derive the time complexity of the recursion using recurrence equations.
For details, see the main paper \cite{recfundam}.

In the unfolder, the complexity of a recursive step mainly depends on the time for copying head and guard, for guard checking, and for unfolding and simplification of the current rule.
In the meta-interpreter, the time complexity is dominated by the third clause that
applies a rule from the list to the current goal.
The complexity of a recursive step in that clause 
is determined by the time needed for copying the rule
and for the meta-calls of the guard and of the two body conjuncts of the rule.
In the round-robin rule processor, the third clause calls the unfolder and the generalized meta-interpreter for each rule list. Note that rule lists are reused in later rounds.

Our complexity estimates are based on the following assumptions that hold in SWI Prolog.
Unification and copying take constant time for given terms and (near-)linear time in the size of the involved terms in general.
A Prolog meta-call has the same time complexity as directly executing its goal argument.
SWI Prolog uses the GNU multiple precision arithmetic library (GMP), where integer arithmetic is unbounded. 
Comparison and addition have logarithmic worst-case time complexity in the numbers involved. 
A variety of multiplication algorithms are used in GMP to get near logarithmic complexity.
If one multiplies with a power of $2$, the complexity is reduced to logarithmic.

In our experiments, we used SWI Prolog 9.2.3-1 running on an
Apple Mac mini M1 2020 MacOS Monterey 12.7.4 with 16 GB RAM.
We use default settings for SWI Prolog 
except for the command line option {\tt -O} which compiles arithmetic expressions.
During multiple runs of the benchmarks we observed a jitter in timings of at most 7\%.
Because the runtime improvement is so dramatic, we can only benchmark small inputs with the original recursion and have to benchmark larger inputs with 
runtime recursion unfolding.

\subsection{Summation Example, Contd.}

We have already unfolded and simplified the recursive rule for summation in Example \ref{sumunf}.
We introduced the implementation in concrete Prolog syntax in Example \ref{sumimpl}.
In \cite{recfundam},
we derived estimates for the time complexities for our summation example written in CHR that we can reuse here when we compare them to benchmark results. 

\subsubsection{Benchmarks}  % Summation Example

Table \ref{benchsum} shows benchmarks results for the summation example.
Times are given in milliseconds.
Since the example features only one recursive rule, there is no need to use the round-robin rule processor {\tt umr/2}.
The benchmarks confirm the super-linear speedup.
\begin{table}[htb] %tb
\begin{center}{%\small
\begin{tabular}{l l l}

\begin{tabular}{|l|r|}
\cline{1-2}
\multicolumn{2}{|c|}{Original Summation} \\
\cline{1-2}
Input $n$  & Time\\
\cline{1-2}
$2^{16}$ & 4 \\
$2^{17}$ & 8 \\
$2^{18}$ & 16 \\
$2^{19}$ & 31 \\
$2^{20}$ & 62 \\
$2^{21}$ & 125 \\
$2^{22}$ & 249 \\
\cline{1-2}
\end{tabular}

& &

\begin{tabular}{|l|r|r|r|}
\cline{1-4}
\multicolumn{4}{|c|}{Runtime Repeated Recursion Unfolding} \\
\cline{1-4}
Input $n$ & Unfolder & Interpreter & Total Time\\
\cline{1-4}
$2^{25}$ & 0.07 & 0.02 & 0.09 \\ 
$2^{50}$ & 0.07 & 0.04 & 0.11 \\
$2^{100}$ & 0.14 & 0.08 & 0.23 \\
$2^{200}$ & 0.29 & 0.17 & 0.46 \\
$2^{400}$ & 0.58 & 0.34 & 0.92 \\
$2^{800}$ & 1.17 & 0.70 & 1.88 \\
$2^{1600}$ & 2.55 & 1.53 & 4.08 \\ 
\cline{1-4}
$2^{25}+1$ & 0.04 & 0.02 & 0.06 \\
$2^{50}+1$ & 0.07 & 0.03 & 0.10 \\
$2^{100}+1$ & 0.15 & 0.06 & 0.21 \\
$2^{200}+1$ & 0.29 & 0.12 & 0.41 \\
$2^{400}+1$ & 0.58 & 0.25 & 0.83 \\
$2^{800}+1$ & 1.20 & 0.50 & 1.70 \\
$2^{1600}+1$ & 2.45 & 1.00 & 3.44 \\
\cline{1-4}
\cline{1-4}
\end{tabular}

\end{tabular}
\caption{Benchmarks for Summation Example (times in milliseconds)}
\label{benchsum}
}\end{center}
\end{table}

\myparagraph{Original Recursion}
In each subsequent table entry, we double the input number.
The runtime roughly doubles.
So the runtime is at least linear.
This is in line with the expected log-linear time complexity $O(n \log(n))$:
since the numbers are so small, addition is fast, almost constant time,
and the runtime is dominated by the linear time overhead of the recursion itself.

\myparagraph{Unfolder and Meta-Interpreter}
For runtime repeated recursion unfolding of our summation example, we give the time needed for the unfolding, the time needed for the execution with the meta-interpreter, and the sum of these timings (column 'Total Time').
Because our method has lower time complexity and is quickly several orders of magnitude faster, we start from $2^{25}$ and in each subsequent table entry, we square the input number instead of just doubling it.

The runtimes of the unfolder and meta-interpreter are similar.
For each squaring of the input number, the their runtimes more than double.
The benchmarks results obtained are consistent with the expected complexities of $O(\log(n)^2)$ of the unfolder and meta-interpreter.

\myparagraph{Comparing Recursion Depths $2^i$ and $2^i+1$}
In the meta-interpreter, each of the unfolded rules will be tried by unifying its head and checking its guard, but not all rules will be necessarily applied.
This may lead to the seemingly counterintuitive behavior that a larger query where only one unfolded rule is applied runs faster than a smaller one where several less unfolded rules are applied.

To see how pronounced this phenomena is,
we compare timings for values of $n$ of the form $2^i$ and $2^i+1$. 
For input numbers of the form $2^i$, all unfolded rules are applied.
Input numbers of the form $2^i+1$, however, will need exactly one application of the most unfolded rule $r_i$ to reach the base case.
As a consequence the runtime of the meta-interpreter decreases by a linear factor of about $0.7$. 
The timings for the unfolder stay about the same, because only one more rule is generated for $2^i+1$ (e.g. $n=2^{1600}+1$ generates 1601 rules).

\subsection{Fibonacci Number Example}

This classical program generates the n-th Fibonacci number in a naive inefficient way using double recursion. 
\begin{verbatim}
fib(N, F) :- N > 1 ,!, 
        N1 is N-1, N2 is N1-1,
        fib(N1, F1), fib(N2, F2),
        F is F1+F2.
fib(N, F) :- N =< 1 ,!, F=N.
\end{verbatim}
The program features double recursion and exponential complexity and therefore cannot be handled by runtime repeated recursion unfolding as defined in \cite{recfundam}. 
With our generalization in the meta-interpreter, it becomes possible. 

\subsubsection{Implementation of Runtime Repeated Recursion Unfolding}

\myparagraph{Rule Template}
This time we start with the desired rule template. It is a generalization of the original recursive rule by keeping the two recursive calls and by introducing parameters in the sum.
\begin{verbatim}
f(N, F) :- N > A ,!, 
        (N1 is N-A, N2 is N1-1),
        (f(N1, F1), f(N2, F2)),
        F is P*F1+Q*F2.
\end{verbatim}
The parameters of the rule template that change during unfolding are {\tt A, P} and {\tt Q}. 
For the original recursive rule, we have that {\tt A=P=Q=1}.

\myparagraph{Unfolding Scheme}
To find the rule template and its unfolding scheme,
the idea was to unfold both recursive calls and
to simplify and to merge them so that again only two recursive calls remain.
The derivation of this unfolding scheme is given in appendix \ref{fibunf}.
The implementation is accomplished by the following clause for {\tt simp\_unf}.
We assume that the input is always a natural number (nonnegative integer).
\begin{verbatim}
simp_unf((f(N, F) :- N > A ,!, 
        (N1 is N-A, N2 is N1-1),
        (f(N1, F1), f(N2, F2)),
        F is P*F1+Q*F2),
        (f(Nc, Fc) :- Nc > Ac ,!, 
        (N1c is Nc-Ac, N2c is N1c-1),
        (f(N1c, F1c), f(N2c, F2c)),
        Fc is Pc*F1c+Qc*F2c)
        ):-
        Ac is 2*A, QQ is Q*Q, Pc is P*P+QQ, Qc is 2*P*Q-QQ.
\end{verbatim}

\myparagraph{Recursive Predicate}
For the Fibonacci example, its rules are replaced by:
\begin{verbatim}
fib(I,O) :- 
    unf(f(I,O), [
       (f(N, F) :- N > 1 ,!, 
         (N1 is N-1, N2 is N1-1), (f(N1, F1),f(N2, F2)), F is 1*F1+1*F2), 
       (f(N, F) :- N=<1 ,!, F=1, true, true)], URs), 
    mip(f(I,O), URs).
\end{verbatim}

\myparagraph{Unfolded Rules}
The first few rules that are returned by the unfolder are
\begin{verbatim}
f(A,D):-A>16,!, (B is A-16,C is B-1), (f(B,E), f(C,F)), D is 1597*E+987*F.
f(A,D):-A>8 ,!, (B is A-8, C is B-1), (f(B,E), f(C,F)), D is 34*E+21*F.
f(A,D):-A>4 ,!, (B is A-4, C is B-1), (f(B,E), f(C,F)), D is 5*E+3*F.
f(A,D):-A>2 ,!, (B is A-2, C is B-1), (f(B,E), f(C,F)), D is 2*E+1*F.
f(A,D):-A>1 ,!, (B is A-1, C is B-1), (f(B,E), f(C,F)), D is 1*E+1*F.
\end{verbatim}

\subsubsection{Complexity}

\myparagraph{Original Recursion}
The recursion depth is determined by the input number $n$. The original recursive rule for Fibonacci gives raise to the 
recurrence equation $f(n)=f(n{-}1)+f(n{-}2)+O(n)$.
$O(n)$ is the complexity of adding the Fibonacci numbers. These numbers grow exponentially with the input, thus their number of bits is linear to the input number.
The solution of the recurrence results is the well-known exponential complexity in $O(2^n)$. 

\myparagraph{Unfolder and Meta-Interpreter}
In the unfolder, when unfolding rule $r_i$, 
the input number $n$ is checked to be larger than parameter {\tt A} in the guard, where {\tt A} can be shown to be $2^i$.
In {\tt simp\_unf/2} the multiplications of the parameters of the rule template dominate
the complexity.
This leads to the recurrence
$u(n) = u(n/2) + M(n)$, 
where the $M(n)$ is the time complexity of efficient multiplication involving a number of $n$ bits.
The solution of the recurrence is in $M(n)$.
The complexity $M(n)$ can be assumed to be $O(n log(n) log(log(n)))$.

In the meta-interpreter, the applications of the unfolded rules determine the complexity.
According to the rule template,
we have two recursive calls and the multiplications with Fibonacci numbers dominate the complexity of a recursive step.
This leads to the recurrence
$f(n) = f(n/2)+f(n/2)+M(n)$.
The solution of the recurrence is in $O(log(n) M(n))$, i.e. $O(n log(n)^2 log(log(n)))$.

Since the example features only one recursive rule, there is no need to use the round-robin rule processor {\tt umr/2}.

\subsubsection{Benchmarks}

Table \ref{benchfib} shows benchmarks results for the Fibonacci example.
Times are in seconds.
A time measurement of 0.0n means that it was below 0.01 but more than zero.
The timings for the original version 
show that the runtime is exponential as expected. 
The runtimes of the unfolder and interpreter roughly double with each doubling of the input number which is consistent with the estimated complexities that are somewhat worse than log-linear in the input number $n$.

\begin{table}[h] %tb
\begin{center}{%\small
\begin{tabular}{l c r}  % nested tabular seems not to work in tplp environment

\begin{tabular}{|l|r|}
\cline{1-2}
\multicolumn{2}{|c|}{Original Fibonacci} \\
\cline{1-2}
Input $n$  & Time\\
\cline{1-2}
$2^5$ & 0n \\
$2^5+1$ & 1 \\
$2^5+2$ & 1 \\
$2^5+3$ & 2 \\
$2^5+4$ & 3 \\
$2^5+5$ & 5 \\
$2^5+6$ & 8 \\
$2^5+7$ & 13 \\
$2^5+8$ & 22 \\
\cline{1-2}
\end{tabular}

& &

\begin{tabular}{|l|r|r|r|}
\cline{1-4}
\multicolumn{4}{|c|}{Runtime Repeated Recursion Unfolding} \\
\cline{1-4}
Input $n$ & Unfolder & Interpreter & Total Time\\
\cline{1-4}
$2^{18}$ & 0,0n & 0.18 & 0.18 \\
$2^{19}$ & 0,0n & 0.34 & 0.34 \\
$2^{20}$ & 0,0n & 0.66 & 0.66 \\
$2^{21}$ & 0.01 & 1.28 & 1.28 \\
$2^{22}$ & 0.02 & 2.48 & 2.50 \\
$2^{23}$ & 0.03 & 4.84 & 4.88 \\
$2^{24}$ & 0.07 & 9.42 & 9.49 \\
\cline{1-4}
$2^{18}+1$ & 0,0n & 0,000n & 0,0n \\
$2^{19}+1$ & 0,0n & 0,000n & 0,0n \\
$2^{20}+1$ & 0.01 & 0.0001 & 0.01 \\
$2^{21}+1$ & 0.02 & 0.0001 & 0.02 \\
$2^{22}+1$ & 0.03 & 0.0001 & 0.04 \\
$2^{23}+1$ & 0.07 & 0.0002 & 0.07 \\
$2^{24}+1$ & 0.15 & 0.0003 & 0.15 \\
\cline{1-4}
\cline{1-4}
\end{tabular}

\end{tabular}
\caption{Benchmarks for Fibonacci Example}
\label{benchfib}
}\end{center}
\end{table}

\myparagraph{Comparing Recursion Depths $2^i$ and $2^i+1$}
For $n$ of the form $2^i$ all unfolded rules are applied in the meta-interpreter. 
The meta-interpreter is two orders of magnitude slower than the unfolder.
The sum of runtimes is therefore dominated by the meta-interpreter.
For $n$ of the form $2^i+1$ on the other hand, only the most unfolded rule is applied in the meta-interpreter.
Its runtime is neglectable because the multiplications involve only the Fibonacci number $1$. 
The sum of runtimes is therefore very much dominated by the unfolder.
The total time is almost two orders of magnitude faster for $2^i+1$ than for $2^i$.

\subsection{Greatest Common Divisor Example}

The Euclidean algorithm is a method for finding the GCD of two numbers by repeatedly subtracting the smaller number from the larger number until both are the same.
The implementation naturally lends itself to using two recursive rules.
\begin{verbatim}
g(M, N, X) :- M<N ,!, L is N-M, g(M, L, X).
g(M, N, X) :- M>N ,!, L is M-N, g(L, N, X).
g(M, M, M).
\end{verbatim}
Since we have multiple recursive rules, we use our extended implementation of generalized runtime repeated recursion unfolding from Section \ref{genimpl} that relies on the predicate {\tt umr/2}.

\subsubsection{Implementation of Runtime Repeated Recursion Unfolding}

\myparagraph{Rule Template}
Note that the two recursive rules have a similar structure and therefore can be unfolded in an analogous way.
Unfolding the rules once with themselves results in the two rules:
\begin{verbatim}
g(M,N,X) :- 2*M<N ,!, L is N-2*M, g(M,L,X).
g(M,N,X) :- M>2*N ,!, L is M-2*N, g(L,N,X).
\end{verbatim}
The generalization is apparent. It is the introduction of a parameter {\tt A} in place of 
the number {\tt 2} with which we multiply the smaller number.
This results in the following template for the two rules:
\begin{verbatim}
g(M,N,X) :- A*M<N ,!, L is N-A*M, g(M,L,X).
g(M,N,X) :- M>A*N ,!, L is M-A*N, g(L,N,X)
\end{verbatim}

\myparagraph{Unfolding Scheme}
From unfolding the rule templates we derive the following two clauses for {\tt simp\_unf},
where parameter {\tt A} is doubled with each unfolding.
\begin{verbatim}
simp_unf((g(M,N,X) :- A*M<N ,!, L is N-A*M, g(M,L,X), true), 
    (g(M1,N1,X1) :- A1*M1<N1 ,!, L1 is N1-A1*M1, g(M1,L1,X1), true)) :-
          A1 is 2*A.
simp_unf((g(M,N,X) :- M>A*N ,!, L is M-A*N, g(L,N,X), true), 
    (g(M1,N1,X1) :- M1>A1*N1 ,!, L1 is M1-A1*N1, g(L1,N1,X1), true)) :-
          A1 is 2*A.
\end{verbatim}

The call is as follows:
\begin{verbatim}
umr(g(A,B,_),[
       [(g(M,N,Z) :- 1*M<N ,!, L is N-1*M, g(M,L,Z), true), 
        (g(M,M,M) :- true ,!, true, true, true)],
       [(g(M,N,Z) :- M>1*N ,!, L is M-1*N, g(L,N,Z), true), 
        (g(M,M,M) :- true ,!, true, true, true)]
       ])
\end{verbatim}

The first few unfolded rules for both recursive clauses are:
\begin{verbatim}
g(M,N,Z) :- 1*M<N ,!, L is N-1*M, g(M,L,Z), true. 
g(M,N,Z) :- 2*M<N ,!, L is N-2*M, g(M,L,Z), true. 
g(M,N,Z) :- 4*M<N ,!, L is N-4*M, g(M,L,Z), true. 
g(M,N,Z) :- 8*M<N ,!, L is N-8*M, g(M,L,Z), true. 

g(M,N,Z) :- M>1*N ,!, L is M-1*N, g(L,N,Z), true. 
g(M,N,Z) :- M>2*N ,!, L is M-2*N, g(L,N,Z), true. 
g(M,N,Z) :- M>4*N ,!, L is M-4*N, g(L,N,Z), true. 
g(M,N,Z) :- M>8*N ,!, L is M-8*N, g(L,N,Z), true. 
\end{verbatim}

\subsubsection{Complexity}

\myparagraph{Original Recursion}
The worst case for the original GCD program is when the input numbers are $n$ and $1$.
Then the repeated subtraction results in a recursion depth of $n$.
Subtraction like addition has logarithmic complexity in the number.
Then the recurrence $g(n) = O(log(n)) + g(n-1)$ leads to the worst-case time complexity
of $O(n \ log(n))$.

\myparagraph{Unfolder, Meta-Interpreter and Round-Robin Processor}
In the unfolder and meta-interpreter, the worst-case time complexity is dominated by the generation and application of the unfolded rules.
In the unfolder, 
the multiplications of the parameters in the rule templates of {\tt simp\_unf/2} dominate the complexity. These multiplications are by a power of $2$ and in this special case have the same complexity as additions.
The recurrence is of the form
$g(n) = O(log(n)) + g(n/2)$
with a resulting  worst-case time complexity
$O(log(n)^2)$.
Similarly, in the meta-interpreter, the multiplications, again by a power of $2$, in the unfolded rules lead to the same complexity.

In the round-robin rule processor {\tt umr/2}, we apply the unfolder and meta-interpreter for several rounds (recursive steps).
Since each round applies at least one rule in the meta-interpreter, a crude upper bound for the number of rounds is $n$ when we alternate between applying the two original rules because no unfolding is possible. 
For a more precise measure, assume a pair of numbers $n$ and $m$ for GCD with $n>m$. 
One round with {\tt umr} will apply one recursive rule and its unfoldings to exhaustion and 
thus will lead to a pair of numbers $n'$ and $m$ where $n'<m$.
Now it holds that $n'< n/2$, 
because we keep subtracting $m$ from $n$ until $n'< m$ and we can subtract at least once, so $n' \leq n-m$. Adding these two inequations gives $2n' < n$ and thus $n'< n/2$.
Hence each round, i.e. each change in rules, at least halves the larger number. Hence the number of rounds is in $O(log(n))$.
Thus the overall worst-case time complexity is in $O(log(n)) O(log(n)^2)$ which is $O(log(n)^3)$.
This is sufficient for a super-linear speedup compared to $O(n \ log(n))$.

\subsubsection{Benchmarks}

Times are in seconds. We discuss two sets of benchmarks.
Table \ref{benchgcd} shows some benchmarks results for the GCD example.
We use powers of $2$ for the larger input number $n$
and a $37$ for the second input number.
For the original rules we use successive exponents up to $33$,
the runtime of the original rules is consistent with the estimated worst-case complexity
$O(n \ log(n))$.
With our generalized runtime repeated recursion unfolding, the runtime 
is so fast that we start from an exponent $5000$ and increase by a factor of about $\sqrt{2}$ up to $40000$.
The runtime is in the estimated complexity of $O(log(n)^3)$, but also consistent with the better
$O(log(n)^2)$. 
Since the second input number is fixed to $37$, after one round in {\tt umr} the first number will be smaller than $37$ and the remaining runtime will be neglectable.

\begin{table}[h] %tb
\begin{center}{%\small
\begin{tabular}{l c r}

\begin{tabular}{|l|r|}
\cline{1-2}
\multicolumn{2}{|c|}{Original GCD} \\
\cline{1-2}
Input $n, 37$  & Time\\
\cline{1-2}
$2^{27}$ & 0.35 \\
$2^{28}$ & 0.64 \\
$2^{29}$ & 1.27 \\
$2^{30}$ & 2.52 \\
$2^{31}$ & 5.04 \\
$2^{32}$ & 10.09 \\
$2^{33}$ & 20.40 \\
\cline{1-2}
\end{tabular}

& &

\begin{tabular}{|l|r|}
\cline{1-2}
\multicolumn{2}{|c|}{Runtime Repeated Recursion Unfolding} \\
\cline{1-2}
Input $n, 37$ & Total Time\\
\cline{1-2}
$2^{5000}$ & 0.015 \\
$2^{7000}$ & 0.021 \\
$2^{10000}$ & 0.032 \\
$2^{14000}$ & 0.045 \\
$2^{20000}$ & 0.067 \\
$2^{28000}$ & 0.094 \\
$2^{40000}$ & 0.148 \\
\cline{1-2}
\end{tabular}

\end{tabular}
\caption{Benchmarks for GCD Example 1}
\label{benchgcd}
}\end{center}
\end{table}

Table \ref{benchgcd2} shows results for another pair of inputs, $n=2^k$ and $m=2^{k//2}+2^{k//4}-1$, i.e. $m$ is somewhat larger than the square root of $n$.
For the original rules we use successive exponents up to $46$,
the runtime is consistent with the estimated worst-case complexity $O(n \ log(n))$.
If we multiply the inputs by about $4$, the runtime roughly doubles.
So the actual complexity for this type of number pairs is in $O(\sqrt{n})$.
With our generalized runtime repeated recursion unfolding, the runtime 
is in the estimated complexity of $O(log(n)^3)$, but also again consistent with the better
$O(log(n)^2)$. 

\begin{table}[h] %tb
\begin{center}{%\small
\begin{tabular}{l c r}

\begin{tabular}{|l|r|}
\cline{1-2}
\multicolumn{2}{|c|}{Original GCD} \\
\cline{1-2}
Input $n$  & Time\\
\cline{1-2}
$2^{40}$ & 0.07 \\
$2^{41}$ & 0.13 \\
$2^{42}$ & 0.13 \\
$2^{43}$ & 0.27 \\
$2^{44}$ & 0.27 \\
$2^{45}$ & 0.53 \\
$2^{46}$ & 0.52 \\
\cline{1-2}
\end{tabular}

& &

\begin{tabular}{|l|r|}
\cline{1-2}
\multicolumn{2}{|c|}{Runtime Repeated Recursion Unfolding} \\
\cline{1-2}
Input $n$ & Total Time\\
\cline{1-2}
$2^{5000}$ & 0.015 \\
$2^{7000}$ & 0.023 \\
$2^{10000}$ & 0.039 \\
$2^{14000}$ & 0.072 \\
$2^{20000}$ & 0.138 \\
$2^{28000}$ & 0.277 \\
$2^{40000}$ & 0.611 \\
\cline{1-2}
\end{tabular}

\end{tabular}
\caption{Benchmarks for GCD Example 2}
\label{benchgcd2}
}\end{center}
\end{table}

\section{Related Work}\label{related}

\emph{Program transformation} to improve efficiency is usually concerned with a strategy for combining unfolding \cite{sterling1994art,Prestwich1993} and folding to replace code
(for an overview see e.g. \cite{visser2005survey,pettorossi2024historical}). 
The transformations are typically performed offline, at compile-time.
{Program transformation} for specific aims and 
applications is abundant in logic programming in general \cite{pettorossi1999synthesis}.

{\em Partial evaluation (partial deduction, program specialization)} \cite{leuschel2002logic} is a program transformation to execute programs with partially known input to specialize it, typically at compile-time.
Compile-time partial evaluation alone cannot achieve super-linear speedup up (Chapter 6 in \cite{Jones-etal93}).
This result does not apply to our approach, because we use a runtime transformation and strongly rely on simplification.
In \cite{bolz2010towards} just-in-time partial evaluation of Prolog is introduced consisting of 1500 lines of Prolog code. A linear speedup of up to a factor of 5 was reported in benchmarks.

In general, {\em super-linear speedups} by program transformation are rare and mostly concern parallel programs.
Our technique applies to sequential programs.
In a sequential setting, 
super-linear speedups can sometimes be achieved with {\em memoization}, where the results of recursive calls are cached and reused if the same recursive call reappears later on.
{\em Tupling} \cite{hu1997} applies when several recursions operate on the same data structure. Then tupling tries to merge these recursions into a single one. 
Then there is work based on \emph{supercompilation} for functional programming languages like Refal and Haskell. 
In advanced forms of this offline program transformation such as distillation \cite{hamilton09} and equality indices \cite{glueck16},
sophisticated generalization while unfolding increases the chance for folding
and can achieve super-linear speedup.

In contrast, our technique relies solely on unfolding and simplifying the combined recursive steps again and again. 
We add redundant rules this way but never remove any. 
Further related work is discussed in the main paper \cite{recfundam}.

\section{Conclusions}

In this companion paper to \cite{recfundam}, we generalized runtime repeated recursion unfolding for multiple recursion and for multiple recursive rules and implemented it in the logic programming language Prolog.
We provided a lean implementation of our approach in ten clauses, comprising the unfolder, a generalized meta-interpreter and a novel round-robin rule processor to handle multiple recursive rules.
We showed with benchmarks for several classical algorithms that the super-linear speedup indeed holds in practice.
We have considered recursive clauses that have green cuts for efficiency. Preliminary experiments show that our implementation also works without such cuts, but the effects on runtime have to be further investigated.
We have not yet covered mutual recursion, but do not see problems in doing so.

Runtime repeated recursion unfolding hinges on finding the appropriate rule template for unfolding for the given problem at compile-time. This requires insight into the given problem and cannot be fully automated.
In the main paper \cite{recfundam},
we discuss these and other issues and limitations of runtime repeated recursion unfolding and suggest some possible improvements as well.

In the main paper
we proved runtime repeated recursion unfolding for single direct recursions correct for CHR 
(Constraint Handling Rules) by showing the redundancy of unfolded recursive rules and their termination.
The proof ideas are expected to carry over to our generalization in Prolog.
In the main paper
we also proved a sufficient and necessary condition for super-linear speedup.
For a given recursion, then one tries to find an unfolding with an improved time complexity that satisfies the condition. If it can be found, a super-linear speedup is guaranteed.
It is left for future work to extend this result to multiple recursion and to Prolog.

We think our approach can also be applied to functional programming languages. For mainstream programming languages it should be possible to adapt the technique to loops as well.
Overall, our generalized runtime repeated recursion unfolding provides a promising strategy for online optimization of recursions in which the sufficient user-definable simplification of combined successive recursive steps leads to predictable speedups that can be super-linear.

\medskip
{\bf Acknowledgements.} 
This research was performed during the sabbatical of the author in the winter semester of 2024/25.
We thank Jesper Larsson Träff for inspiring discussions.

\bibliographystyle{alpha} %! abbrv, alpha is longer
\bibliography{recunfold,biblio,chrjust}

\appendix

\section{Fibonacci Numbers, Derivation of Rule Scheme for Unfolding}\label{fibunf}

To find the unfolding scheme for {\tt simp\_unf/2},
the idea is to unfold both recursive calls,
to simplify and merge them so that again only two recursive calls remain.
Recall the rule template for Fibonacci numbers
\begin{verbatim}
f(N, F) :- N > A ,!, 
        (N1 is N-A, N2 is N1-1),
        (f(N1, F1), f(N2, F2)),
        F is P*F1+Q*F2.
\end{verbatim}
In the rule template, we first unfold both recursive calls and 
collect the guards together with the relevant arithmetic computations.
\begin{verbatim}
f(N, F) :- N > A, (N1 is N-A, N2 is N1-1), N1 > A, N2 > A ,!, 
        % remaining unfolding of f(N1, F1)
        (N11 is N1-A, N21 is N11-1),
        (f(N11, F11), f(N21, F21)),
        F1 is P*F11+Q*F21,
        % remaining unfolding of f(N2, F2),
        (N12 is N2-A, N22 is N12-1),
        (f(N12, F12), f(N22, F22)),
        F2 is P*F12+Q*F22,
        %
        F is P*F1+Q*F2.
\end{verbatim}

According to the original rule template, we want to have just two recursive calls. 
Consider the variables {\tt N12} and {\tt N21}.
They depend on {\tt N1} in the following way: {\tt N12 is N2-A, N2 is N1-1}
and
{\tt N21 is N11-1, N11 is N1-A}.
We observe that this implies that {\tt N21=N12} and therefore {\tt F21=F12}, since Fibonacci is a function.
Hence we do not need to compute {\tt f(N12, F12)} and can remove it.
\begin{verbatim}
f(N, F) :- N > A, (N1 is N-A, N2 is N1-1), N1 > A, N2 > A ,!, 
        % remaining unfolding of f(N1, F1)
        (N11 is N1-A, N21 is N11-1),
        (f(N11, F11), f(N21, F21)),
        F1 is P*F11+Q*F21,
        % remaining unfolding of f(N2, F2),
        (N12 is N2-A, N22 is N12-1),
        (F12=F21, f(N22, F22)),
        F2 is P*F12+Q*F22,
        %
        F is P*F1+Q*F2.
\end{verbatim}
This leaves three recursive calls with successive input numbers in decreasing order,
{\tt f(N11, F11), f(N21, F21), f(N22, F22)}.
By definition of Fibonacci numbers, we have that {\tt F11 is F21+F22}.
We can use this equation to compute {\tt F22} by {\tt F22 is F11-F21}
instead of computing it with the third recursive call {\tt f(N22, F22)}.
\begin{verbatim}
f(N, F) :- N > A, (N1 is N-A, N2 is N1-1), N1 > A, N2 > A ,!, 
        % remaining unfolding of f(N1, F1)
        (N11 is N1-A, N21 is N11-1),
        (f(N11, F11), f(N21, F21)),
        F1 is P*F11+Q*F21,
        % remaining unfolding of f(N2, F2),
        (N12 is N2-A, N22 is N12-1),
        (F12=F21, F22 is F11-F21),
        F2 is P*F12+Q*F22,
        %
        F is P*F1+Q*F2.
\end{verbatim}
We have successfully removed the last two recursive calls
from the unfolding of {\tt f(N2,F2)}. 

Now we simplify the arithmetic expressions, first by removing variables
that have become superfluous.
The variables {\tt N12, N22} 
occurring in {\tt N12 is N2-A, N22 is N12-1} are no longer needed.
The condition {\tt N2 > A} that remained from the unfolding of {\tt f(N2,F2)}
is no longer needed
and thus also {\tt N2} can be removed.
\begin{verbatim}
f(N, F) :- N > A, (N1 is N-A), N1 > A ,!, 
        % remaining unfolding of f(N1, F1)
        (N11 is N1-A, N21 is N11-1),
        (f(N11, F11), f(N21, F21)),
        F1 is P*F11+Q*F21,
        % remaining unfolding of f(N2, F2),
        (F12=F21, F22 is F11-F21),
        F2 is P*F12+Q*F22,
        %
        F is P*F1+Q*F2.
\end{verbatim}
We continue simplification by removing the intermediate variables {\tt N1, F1, F2, F12} and {\tt F22}.
\begin{verbatim}
f(N, F) :- N > A, N-A > A ,!, 
        % remaining unfolding of f(N1, F1)
        (N11 is N-A-A, N21 is N11-1),
        (f(N11, F11), f(N21, F21)),
        %
        F is P*(P*F11+Q*F21)+Q*(P*F21+Q*(F11-F21)).
\end{verbatim}
We further simplify the remaining arithmetic expressions to arrive finally at:
\begin{verbatim}
f(N, F) :- N > 2*A ,!, 
        % remaining unfolding of f(N1, F1)
        (N11 is N-2*A, N21 is N11-1),
        (f(N11, F11), f(N21, F21)),
        %
        F is (P*P+Q*Q)*F11+(2*P*Q-Q*Q)*F21.
\end{verbatim}
This rule fits the rule template that we have initially devised.

\section{Further Examples}

The examples for list reversal and sorting are taken from the main paper \cite{recfundam}, where the details can be found such as the derivation of the unfolding scheme and of the time complexity results.
We did a straightforward translation from CHR \cite{rpbook} to Prolog and performed the benchmarks. 
They show the expected super-linear speedups.

\subsection{List Reversal Example}

The classical program reverses a given list in a naive way.
It takes the first element of the list, reverses its remainder and adds the element to the end of the reversed list. 
The Prolog predicate {\tt r(A,B)} holds if list {\tt B} is the reversal of list {\tt A}.
\begin{verbatim}
r(E, D) :- E=[C|A], !, r(A, B), append(B, [C], D).
r(E, D) :- E=[], !, D=[].
\end{verbatim}
The predicate {\tt append(X,Y,Z)} concatenates two lists {\tt X} and {\tt Y} into a third list {\tt Z}. Its runtime is linear in the length (number of elements) of the first list.

\subsubsection{Implementation of Runtime Repeated Recursion Unfolding}

\myparagraph{Unfolding with Simplification}
The unfolding scheme for list reversal is implemented with the following clause for {\tt simp\_unf/2}. 
\begin{verbatim}
simp_unf(
  (r(A,B) :- A=E ,!, true, r(C,D), append(D,F,B)), % given rule template
  (r(Al,Bl) :- Al=El ,!, true, r(Cl,Dl), append(Dl,Fl,Bl)) % unfolded r.
  ) :-
    copy_term((E,C,F),(El,Cc,Fc1)), 
    copy_term((E,C,F),(Ec,Cl,Fc2)), 
    Cc=Ec,
    append(Fc2,Fc1,Fl).  
\end{verbatim}
During unfolding,
in the given rule template,
the variable {\tt E} in the guard will be instantiated with an open list ending in the variable {\tt C}.
The list {\tt F} in {\tt append/3} then consists of the element variables of {\tt E} in reversed order.
In the unfolded rule template, the number of elements in these two lists is doubled and their relationship of reversal is maintained.

\myparagraph{Recursive Predicate}
For list reversal, its rules are replaced by:
\begin{verbatim}
rev(I,O) :- 
    unf(r(I,O), [
       (r(A, E) :- A=[D|B] ,!, true, r(B, C), append(C, [D], E), 
       (r(A, B) :- A=[] ,!, B=[], true, true)
                ], URs), 
    mip(r(I,O), URs).
\end{verbatim}
The list in the second argument of {\tt unf/3} contains the original recursive rule and the rule for the base case in appropriate template form.

\myparagraph{Unfolded Rules}
The rules that are returned by the unfolder {\tt unf/3} for a query 
with 17 list elements are
\begin{verbatim}
r(A, T) :- A=[S,R,Q,P,O,N,M,L,K,J,I,H,G,F,E,D|B] ,!, 
        true, r(B, C), append(C, [D,E,F,G,H,I,J,K,L,M,N,O,P,Q,R,S], T).
r(A, L) :- A=[K,J,I,H,G,F,E,D|B] ,!, 
        true, r(B, C), append(C, [D,E,F,G,H,I,J,K], L).
r(A, H) :- A=[G,F,E,D|B] ,!, true, r(B, C), append(C, [D,E,F,G], H).
r(A, F) :- A=[E,D|B] ,!, true, r(B, C), append(C, [D,E], F).
r(A, E) :- A=[D|B] ,!, true, r(B, C), append(C, [D], E).
r(A, B) :- A=[] ,!, B=[], true, true.
\end{verbatim}
The number of element variables in the lists doubles with each unfolding but will never exceed $n$, the length of the input list in the query.
Hence the unfolded rules do not increase overall space complexity.

\subsubsection{Benchmarks}

Table \ref{benchrev} shows benchmarks results for the list reversal example.
The list sizes (lengths) $n$ are powers of $2$.
Times are in seconds.
The experiments confirm the super-linear speedup using runtime repeated recursion unfolding. 

\begin{table}[h] %tb
\begin{center}{%\small
\begin{tabular}{l c l}

\begin{tabular}{|l|r|}
\cline{1-2}
\multicolumn{2}{|c|}{Original list reversal} \\
\cline{1-2}
Input $n$  & Time\\
\cline{1-2}
$2^{9}$ & 0.02 \\
$2^{10}$ & 0.04 \\
$2^{11}$ & 0.09 \\
$2^{12}$ & 0.31 \\
$2^{13}$ & 1.21 \\
$2^{14}$ & 4.81 \\
$2^{15}$ & 19.31 \\
\cline{1-2}
\end{tabular}

& &

\begin{tabular}{|l|r|r|r|}
\cline{1-4}
\multicolumn{4}{|c|}{Runtime Repeated Recursion Unfolding} \\
\cline{1-4}
Input $n$ & Unfolder & Interpreter & Total Time\\
\cline{1-4}
$2^{13}-1$ & 0.0n & 0.0n & 0.0n \\
$2^{14}-1$ & 0.0n & 0.0n & 0.01 \\
$2^{15}-1$ & 0.01 & 0.01 & 0.02 \\
$2^{16}-1$ & 0.02 & 0.01 & 0.03 \\
$2^{17}-1$ & 0.03 & 0.02 & 0.06 \\
$2^{18}-1$ & 0.07 & 0.05 & 0.12 \\
$2^{19}-1$ & 0.13 & 0.10 & 0.23 \\
\cline{1-4}
$2^{13}$ & 0.0n & 0.0n & 0.01 \\
$2^{14}$ & 0.01 & 0.0n & 0.01 \\
$2^{15}$ & 0.02 & 0.01 & 0.03 \\
$2^{16}$ & 0.03 & 0.02 & 0.05 \\
$2^{17}$ & 0.07 & 0.04 & 0.11 \\
$2^{18}$ & 0.13 & 0.08 & 0.21 \\
$2^{19}$ & 0.26 & 0.15 & 0.42 \\
\cline{1-4}
\cline{1-4}
\end{tabular}

\end{tabular}
\caption{Benchmarks for List Reversal Example}
\label{benchrev}
}\end{center}
\end{table}

\myparagraph{Original Recursion}
For the original recursion,
the benchmarks indicate a complexity consistent with the expected $O(n^2)$.
From $2^{12}$ on,
doubling the list size increases the runtime by a factor of about four.

\myparagraph{Unfolder and Meta-Interpreter}
All measured runtimes are consistent with a linear complexity $O(n)$. 
For list size $n = 2^{13}$, runtime repeated recursion unfolding is already 
two orders of magnitude faster than the original recursion.

\myparagraph{Comparing Recursion Depths $2^i-1$ and $2^i$}
We give timings for list lengths $n$ of the form $2^i$ and their predecessor numbers $2^i-1$. 
In the meta-interpreter,
the runtime of applying {\em all} unfolded rules (case of $n=2^i-1$) 
is less than of applying just the next larger unfolded rule 
(which has twice the size) 
(case of $n=2^i$).
The unfolder takes longer than the meta-interpreter.
Going from $2^{i-1}$ to $2^i$, the unfolder generates one more rule and 
the time spent roughly doubles.

\subsection{Sorting Example}

The classical insertion sort program sorts the numbers given in a list in ascending order:
\begin{verbatim}
s(L,S) :- L=[A|L1] ,!, s(L1,S1), i(A,S1,S).
s([],S):- !, S=[].
\end{verbatim}
The predicate {\tt i(A, S1, S)} inserts a number {\tt A} into the sorted list {\tt S1} such that the resulting list {\tt S} is sorted.

\subsubsection{Implementation of Runtime Repeated Recursion Unfolding}

In the rule template that we derive in \cite{recfundam}, 
we use a built-in for merging {\tt m(S1,S2,S3)} instead of insertions.
It merges two sorted lists into a third sorted list.

\myparagraph{Unfolding with Simplification}
Relying on the rule template, 
the unfolding scheme is defined the following clause for {\tt simp\_unf}.
\begin{verbatim}
simp_unf(
   Rule,                                         % given rule
   (s(L,S) :- L=AL ,!, MG, s(L2,S2), m(S0,S2,S)) % unfolded rule template
        ):-     
    copy_term(Rule, (s(L,S) :- L=AL ,!, MG1, s(L1,S1), m(S3,S1,S))), 
    copy_term(Rule, (s(L1,S1) :- L1=AL1 ,!, MG2, s(L2,S2), m(S4,S2,S1))),
    clean((MG1, MG2, m(S3,S4,S0)), MG),
    L1=AL1.
\end{verbatim}
We copy the input rule twice onto instances of the rule template to simulate the unfolding of the recursive call.
The predicate {\tt clean/2} removes superfluous {\tt true} built-ins in the resulting mergings.
Finally, executing {\tt L1=AL1} will double the size of the open list {\tt AL} which ends in {\tt L1}.

\myparagraph{Recursive Predicate}
For the sorting example, its rules are replaced by:
\begin{verbatim}
sort(I,O) :- 
    unf(s(I,O), [
       (s(A,E) :- A=[C|B] ,!, true, s(B,D), m([C],D,E)), 
       (s([],A):- true ,!, A=[], true, true)
                ], URs), 
    mip(s(I,O), URs).
\end{verbatim}
We write the original recursive clause according to the rule template using merge {\tt m/3} instead of insert {\tt i/3}.

\myparagraph{Unfolded Rules}
The first few rules that are returned by the unfolder are
\begin{verbatim}
s(A,S) :- A=[C,B,E,D,I,H,K,J|P] ,!, 
         ((m([B],[C],G), m([D],[E],F), m(F,G,O)), 
          (m([H],[I],M), m([J],[K],L), m(L,M,N)), m(N,O,Q)), 
           s(P,R), m(Q,R,S).
s(A,K) :- A=[C,B,E,D|H] ,!, 
          (m([B],[C],G), m([D],[E],F), m(F,G,I)), 
           s(H,J), m(I,J,K).
s(A,G) :- A=[C,B|D] ,!, m([B],[C],E), s(D,F), m(E,F,G).
s(A,E) :- A=[C|B] ,!, true, s(B,D), m([C],D,E).
s([],A):- true ,!, A=[], true, true.
\end{verbatim}
As with list reversal, the rule size roughly doubles with each unfolding, but again this does not increase the space complexity.

\subsubsection{Benchmarks}

Table \ref{benchsort} shows benchmarks results for the sorting example.
Times are in seconds.
The benchmarks are performed with random permutations of integers from $1$ to $n$.
The experiments confirm the super-linear speedup.

\begin{table}[h] %tb
\begin{center}{%\small
\begin{tabular}{l c l}

\begin{tabular}{|l|r|}
\cline{1-2}
\multicolumn{2}{|c|}{Original Sorting} \\
\cline{1-2}
Input $n$  & Time\\
\cline{1-2}
$2^{8}$ & 0.0n \\ 
$2^{9}$ & 0.02 \\
$2^{10}$ & 0.05 \\
$2^{11}$ & 0.13 \\
$2^{12}$ & 0.48 \\
$2^{13}$ & 1.93 \\
$2^{14}$ & 7.77 \\
\cline{1-2}
\end{tabular}

& &

\begin{tabular}{|l|r|r|r|}
\cline{1-4}
\multicolumn{4}{|c|}{Runtime Repeated Recursion Unfolding} \\
\cline{1-4}
Input $n$ & Unfolder & Interpreter & Total Time\\
\cline{1-4}
$2^{12}-1$ & 0.01 & 0.02 & 0.03 \\
$2^{13}-1$ & 0.01 & 0.03 & 0.03 \\
$2^{14}-1$ & 0.01 & 0.04 & 0.04 \\
$2^{15}-1$ & 0.01 & 0.07 & 0.08 \\
$2^{16}-1$ & 0.02 & 0.15 & 0.17 \\
$2^{17}-1$ & 0.05 & 0.30 & 0.34 \\
$2^{18}-1$ & 0.09 & 0.64 & 0.74 \\
\cline{1-4}
$2^{12}$ & 0.0n & 0.01 & 0.01 \\
$2^{13}$ & 0.01 & 0.02 & 0.02 \\
$2^{14}$ & 0.01 & 0.03 & 0.05 \\
$2^{15}$ & 0.02 & 0.07 & 0.09 \\
$2^{16}$ & 0.05 & 0.14 & 0.18 \\
$2^{17}$ & 0.09 & 0.29 & 0.38 \\
$2^{18}$ & 0.18 & 0.60 & 0.78 \\
\cline{1-4}
\cline{1-4}
\end{tabular}

\end{tabular}
\caption{Benchmarks for Sorting Example}
\label{benchsort}
}\end{center}
\end{table}

\myparagraph{Original Recursion}
The experiments for the original version of insertion sort indicate a complexity that is indeed quadratic $O(n^2)$.
Doubling the list length increases the runtime by a factor of four
from $n=2^{11}$ on.

\myparagraph{Unfolder and Meta-Interpreter}
All timings are linear in $n$ or slightly worse.
The runtimes of the unfolder are consistent with the expected linear complexity $O(n)$.
The meta-interpreter timings are consistent with the expected log-linear complexity $O(n \log(n))$. 
The generation of all rules in the unfolder takes a fraction of the time of applying one or more rules in the meta-interpreter.

\myparagraph{Comparing Recursion Depths $2^i-1$ and $2^i$}
Going from input list length $2^i-1$ to $2^i$, 
the unfolder generates one more rule. 
It has twice the size of the previous rule.
And indeed the runtime for the unfolder roughly doubles.
Going from list length $2^i-1$ to $2^i$, 
the meta-interpreter applies all unfolded rules in the first case but only the next more unfolded rule in the second case. 
Its runtime decreases somewhat.
But overall, the total runtime increases somewhat.

\end{document}